\documentstyle[twocolumn,aps]{revtex}
\newcommand{\be}{\begin{equation}}
\newcommand{\ee}{\end{equation}}
\newcommand{\bea}{\begin{eqnarray}}
\newcommand{\eea}{\end{eqnarray}}
\begin{document}
\bibliographystyle{prsty}
\title{
Bound Pairs of Fronts in a Real Ginzburg-Landau Equation Coupled
to a Mean Field}
\author{Henar Herrero}
\address{Departamento de F\'{\i}sica y Matem\'atica Aplicada,
  Facultad de Ciencias\\
Universidad de Navarra, 31080 Pamplona, Navarra, Spain}
\author{Hermann Riecke}
\address{
Department of Engineering Sciences and Applied Mathematics \\
Northwestern University, Evanston, IL 60208, USA
}
\maketitle
\begin{abstract}
Motivated by the observation of localized traveling-wave states (`pulses')
 in convection in binary liquid mixtures,
the interaction of fronts is investigated in a real Ginzburg-Landau equation
which is
coupled to a mean field. In that system the Ginzburg-Landau equation describes
the traveling-wave amplitude and the
mean field corrsponds to a concentration mode which arises due to the
slowness of mass diffusion. For
single fronts the mean field can lead to a hysteretic transition between slow
and
fast fronts.
Its contribution to the interaction between fronts can be attractive as well
as repulsive and depends strongly on their direction of propagation. Thus,
the concentration mode leads to a new localization mechanism, which does
not require any dispersion in contrast to that operating in the nonlinear
Schr\"odinger equation. Based on this mechanism
alone, pairs of fronts in binary-mixture convection are expected
to form {\it stable} pulses if
they travel {\it backward}, i.e. opposite to the phase velocity. For positive
velocities
the interaction becomes attractive and destabilizes the pulses. These results
are
 in qualitative agreement with recent experiments. Since the new mechanism is
very
robust it is expected to be relevant in other systems as well in which a wave
is
coupled to a mean field.

\end{abstract}


submitted to Physica D, August 8, 1994

\section{Introduction}

Among the most striking recent findings in the investigation of pattern-forming
dissipative systems have been the observations of stable {\it localized}
structures
surrounded either by the structureless
basic state or by other structures. Steady structures of this kind have
been found in experiments on Rayleigh-B\'enard convection in narrow
channels where patches of rolls with long wavelength coexist with
patches of short wavelength \cite{HeVi92,RaRi94,RaRi94b}. In Taylor vortex flow
large, turbulent axisymmetric vortices have been found to coexist stably
with small, laminar vortices \cite{HeAn89}. Another class of localized
structures is given by the localized drift waves observed in a variety
of systems undergoing a parity-breaking bifurcation. They were first found
in directional solidification \cite{SiBe88}. Subsequently they
were also observed in other interface instabilities like the printer
instability \cite{RaMi90} and in cellular flames \cite{GoHa93}. Based on
theoretical predictions \cite{RiPa92},
the same phenomena were then found in Taylor vortex flow between
counter-rotating cylinders \cite{WiAl92}. Theoretically, these localized drift
waves have been successfully described with phase-amplitude equations
adequate for a parity-breaking bifurcation
\cite{RiPa92,CoGo89,GoGu90,GoGu91,CaCa92,BaMa92,BaMa94,BaMa94a}.

The present paper is motivated by localized waves which have been
observed in the convection of binary mixtures. They consist of
traveling-wave packets (`pulses') which drift through the conductive,
structureless
state.
They have been investigated in great detail and presumably constitute the
 best studied
structures of this kind \cite{MoFi87,HeAh87,KoBe88,Ko91,Ko94}.
Depending on parameters, two seemingly different classes of pulses
have been found. For weakly negative separation ratio $\psi$ of the mixture
the pulses
comprise only a few wavelengths and their size is relatively independent
of the parameters (pulses of `fixed width'). For more negative values of the
separation
ratio long packets have been observed. In early experiments their width
varied from run to run in a seemingly random manner (pulses of
`arbitary width'). In those experiments both types of pulses were
stationary,
with the convection rolls traveling through the pulse. In numerical simulations
of the
Navier-Stokes equations \cite{BaLu91,LuBa92} and in subsequent improved
experiments it was found that the short pulses \cite{Ko91} and then
also the long pulses \cite{Ko93} in fact drift as had been expected
on general symmetry grounds. The drift velocity
was, however,  found to be extremely small.

Analytically, these pulses have been described
within the framework of a complex Ginzburg-Landau equation. Two approaches have
been taken.
In the strongly dispersive limit the Ginzburg-Landau equation can be viewed as
a dissipatively perturbed nonlinear Schr\"odinger equation and it has been
shown
that the pulses can arise from the solitons of that equation
\cite{ThFa88,SaHO92,Pi87,MaNe91}.
For weak dispersion localized
waves have been described by fronts which can form stably bound pairs due to
the repulsive
interaction arising from the dispersion \cite{MaNe90,HaPo91}. In either case,
the
localized waves should travel with a velocity which is essentially given by the
linear group
velocity of the waves. The experimentally and numerically observed velocity is,
however, by a factor of
20 to 30 smaller \cite{Ko91,BaLu91}. In addition, the numerical simulations
revealed a striking behavior of
 the concentration field, which is advected by
the traveling rolls. It was suggested that it could be responsible for the
slow-down of the pulses \cite{BaLu91}.
This finding motivated a detailed analysis of the derivation of the
conventional
Ginzburg-Landau equations. It was shown that these equations break down already
quite close
to threshold; due to the slow mass diffusion
in liquids, which is characterized by their small Lewis number
${\cal L} =O(10^{-2})$, certain modes of the concentration decay very slowly
and have
to be treated as independent dynamical variables in addition to the convective
amplitude $A$ \cite{Ri92,Ri92a,Ri92b}. For free-slip-permeable boundary
conditions extended Ginzburg-Landau equations for $A$ and a large-scale
concentration mode $C$ have
been derived. In contrast to the conventional Ginzburg-Landau equation,
they do not become singular in the limit ${\cal L} \rightarrow 0$
and constitute a minimal model to describe this important physical effect.

Numerical simulations of the extended Ginzburg-Landau equations in the strongly
dispersive case showed that the
concentration field can indeed slow down the pulses considerably
\cite{Ri92,Ri92a,Ri92b}.
Analytical investigations using soliton perturbation theory confirm this result
\cite{Ri92b,Riunpub}. Numerically, it has also been found that (long)
pulses can arise stably even {\it without dispersion} when the concentration
mode is included
\cite{Ri92,Ri92a}. This shows, that in this system a second mechanism, which is
unrelated
to that responsible for the solitons of the nonlinear Schr\"odinger equation,
may be able to
 induce the localization
of waves. In the present communication we discuss long pulses considering them
as a bound
pair of fronts. We isolate their localization mechanism analytically and show
that the interaction between fronts can become {\it repulsive} if the pulse
travels
{\it backward}, i.e. opposite to the velocity of the waves. This mechanism is
quite robust and may
be expected to be relevant in other systems as well. Equations similar to those
discussed here have been derived, for instance, for two-fluid channel flow
\cite{ReRe94}.

Recently, the interaction of fronts in
a system of reaction-diffusion type with two fields has been investigated
\cite{HaMe92}. In that work stably bound pairs of fronts have been found as
well. Their localization
mechanism is, however, quite different from
that discussed here; while in those equations the coupling of the two fields
arises through reaction terms, it involves gradient terms in the extended
Ginzburg-Landau
equations discussed here.

The organization of the paper is as follows. In sec.II the extended
Ginburg-Landau equations are discussed. The equations describing the
interaction of fronts are derived in sec.III.
They are analyzed in detail in sec.IV. In sec.V the effect of terms neglected
in the derivation of the equations describing the interaction of the fronts is
discussed
using numerical simulations of the
extended Ginzburg-Landau equations. In the concluding sec.VI we comment on the
relevance of
the concentration mode in combination with dispersion for understanding very
recent experiments
\cite{Ko94}.

\section{The Extended Ginzburg-Landau Equations}

As shown in \cite{Ri92,Ri92a}  the derivation of the conventional
Ginzburg-Landau
equations from the Navier-Stokes equations becomes invalid in the limit
of vanishing mass diffusion of the second component. Since the Lewis number
${\cal L}$, which
is given by the ratio of mass diffusion to heat diffusion, is $O(10^{-2})$ in
liquids this implies that
the conventional Ginzburg-Landau equation yields a poor description of the
experimental system
already for quite small convection amplitudes.
 Technically, in the limit ${\cal L} \rightarrow 0$ a certain concentration
mode $C$ cannot be eliminated adiabatically in favor of the
traveling-wave amplitude $A$ and has to be kept
as an independent dynamical variable. For free-slip-permeable boundary
conditions
this leads to the following equations
as a minimal model to capture the dynamics of the concentration mode
\cite{Ri92},
\bea
\partial_{\tilde{t}} \tilde{A} + \tilde{s} \partial_{\tilde{x}}\tilde{A}
&=&\tilde{d} \partial_{\tilde{x}}^2 \tilde{A} + \tilde{a} \tilde{A} + \tilde{c}
|\tilde{A}|^2 \tilde{A}+
\tilde{p} |\tilde{A}|^4 \tilde{A}+ \tilde{f} \tilde{C} \tilde{A},
\label{e:Atil}\\
\partial_{\tilde{t}} \tilde{C} &=& \tilde{d}_c \partial_{\tilde{x}}^2 \tilde{C}
+\tilde{a}_c \tilde{C} + \tilde{h} \partial_{\tilde{x}} |\tilde{A}|^2 +
\tilde{g} \tilde{C} |\tilde{A}|^2,
\label{e:Ctil}
\eea
where the amplitude $\tilde{A}$ denotes the traveling-wave amplitude and is
identical to that
appearing in the conventional Ginzburg-Landau equation. The streamfunction
$\tilde{\psi}$ and the deviations from the linear temperature profile as well
as
that from the concentration
profile are therefore given by
$(\tilde{\psi},\tilde{\theta},\tilde{\eta})=\epsilon \tilde{A}
 e^{iq\hat{x}+i\omega \hat{t}} \sin \pi \hat{z} (\tilde{\psi_0},
\tilde{\theta_0}, \tilde{\eta_0})$.
The new concentration mode $\tilde{C}$
characterizes a long-wavelength mode of the concentration field which
depends only on the vertical coordinate $\hat{z}$,
$(\tilde{\psi},\tilde{\theta},\tilde{\eta})=\epsilon C \sin2\pi \hat{z}
(0,0,1)$.
{}From a more general perspective, eqs.(\ref{e:Atil},\ref{e:Ctil}) describe the
interaction
of a finite-wavenumber traveling wave with a zero-wavenumber steady mode.
The interaction of two steady
modes with very different wave numbers has been treated in \cite{MePr92}.

To be precise, eqs.(\ref{e:Atil},\ref{e:Ctil})
do not contain all terms up to fifth order. The complete set up to third order
including
the values of the coefficients\footnote{In \cite{Ri92a} it had been overlooked
that the coefficient $h_4$ has to vanish in general due to the fact that the
concentration mode $C$ is real \cite{KoHopriv}.} for free-slip-permeable
boundary conditions
 have been presented in \cite{Ri92a}. Here we want to focus on the advection
of $C$ by the waves and the resulting feed-back on the traveling-wave amplitude
through
the dependence of its growth rate on the local (vertical) gradient in the
concentration
field. These important effects are described in
eqs.(\ref{e:Atil},\ref{e:Ctil}) by the terms
$\tilde{h} \partial_{\tilde{x}} |\tilde{A}|^2$ and
$\tilde{f} \tilde{C} \tilde{A}$, respectively. In addition, the damping term
$\tilde{a}_c\tilde{C}$ ($\tilde{a}_c <0$) arises from vertical diffusion of the
concentration.
Through the term $C |A|^2$ this damping is also affected by convection.
Note that eqs.(\ref{e:Atil},\ref{e:Ctil}) are only adequate if the convective
amplitude is small enough to satisfy $\tilde{a}_c+\tilde{g}|\tilde{A}|^2 < 0$.
If this term were positive, plain waves $|A|^2=A_0^2$
would be unstable to an evergrowing concentration mode.
The leading-order contribution to the diffusive term $\partial_{\tilde{x}}^2C$
does not arise from
diffusion of the concentration; instead large-scale variations
in the concentration field affect the  local bouyancy of the fluid and
generate vorticity which
in turn advects the basic (linear) concentration profile.

To focus on the new localization mechanism, which was found numerically not to
rely
on the dispersive effects embodied in the imaginary parts
of the coefficients \cite{Ri92a}, we consider
here the special case in which all
the coefficients and the amplitude $A$ are real. In addition, to study the
dynamics and interaction
of fronts we restrict ourselves to the regime below threshold, $\tilde{a} < 0$.
Eqs.(\ref{e:Atil},\ref{e:Ctil}) can then be simplified by introducing
scaled variables {\it via}
\bea
A=\left(\frac{\tilde{p}}{\tilde{a}}\right)^{1/4}\tilde{A}, \ \ \
C=-\frac{\tilde{f}}{\tilde{a}}\tilde{C}, \label{e:scal1}\\
t=-\tilde{a} \tilde{t}, \ \ \ x= \sqrt{\frac{-\tilde{a}}{\hat{d}}} \tilde{x}, \
\ \  \tilde{d}=\eta^2 \hat{d}, \label{e:scal2}\\
s=\frac{\tilde{s}}{\sqrt{-\tilde{a}\hat{d}}},\ \ \
c=\frac{\tilde{c}}{\sqrt{\tilde{a}\tilde{p}}}, \ \ \ \delta =
\frac{\hat{d}_c}{\tilde{d}}, \label{e:scal3} \\
 \alpha =  \frac{\tilde{a}_c}{\tilde{a}},\ \ \
h = \frac{\tilde{h}\tilde{f}}{\sqrt{-\tilde{p}\tilde{a}^2\hat{d}}}, \ \ \ g =
\frac{\tilde{g}}{\sqrt{\tilde{a}\tilde{p}}}.\label{e:scal4}
\eea
This leads to
\bea
\partial_tA+ s \partial_xA&=&\eta^2\partial_x^2A  -A+
cA^3-A^5+CA,\label{e:caa}\\
\partial_tC&=& \delta\partial_x^2C-\alpha C+ h\partial_xA^2 + g C
A^2.\label{e:cac}
\eea
Note that in the scaling (\ref{e:scal1}-\ref{e:scal4}) the reduced
Rayleigh number $\tilde{a}$ has been scaled to -1. The
parameter which allows a scanning between the Hopf bifurcation
at $\tilde{a}=0$ and the saddle-node bifurcation
at $\tilde{a}=\tilde{c}^2/4\tilde{p}$ is now
given by $c$.
Its values range therefore  from $c=2$ at the saddle node to $c \rightarrow
\infty$ at the Hopf bifurcation. For free-slip-permeable boundary conditions
the coefficients
$\alpha$, $\delta$, $g$ and, in particular, $h$ are positive \cite{Ri92a}. In
(\ref{e:scal2}) we have introduced
the parameter $\eta$.
It characterizes the strength of the diffusion of the traveling-wave amplitude
$A$ and therefore governs the width of the fronts in $A$.
 In this paper we will consider the limit of
narrow fronts, i.e. $\eta \ll 1$. This allows the derivation
of equations describing the dynamics and interaction of fronts which focusses
on the
contribution from the concentration mode.

\section{Derivation of the Front Equations}

The goal of this paper is an analytical description of the contribution
of the concentration mode to the interaction of fronts. In the standard
approach to capture the interaction of fronts analytically one assumes
the distance between the fronts to be large. The interaction, which
is essentially due to the overlap of the fronts in the convective
amplitude, is then weak and the resulting
dynamics slow. In the present system the fronts interact additionally
{\it via} the concentration mode. In certain parameter regimes the  decay
length of the concentration mode can be much larger than that of the convective
amplitude. The
interaction is then dominated by the concentration mode. Here we describe this
regime
by considering the limit of weak diffusion of $A$, $\eta \ll 1$.
The fronts in $A$ are then steep and the overlap between them can be
ignored for any distance of O(1).

 Based on
numerical simulations \cite{Ri92,Ri92a} and a preliminary calculation
for the interaction between widely separated fronts ($\eta = O(1)$)
\cite{HeRiunpub}
we expect that stable pulses exist only if their
velocity is opposite to the group velocity $s$ (for $h > 0$). Since the
difference between these two velocities has to be small in a perturbative
approach we are led
to assume, in addition, that the group velocity $s$ is small, $s=\eta^2 s_2$.
Consequently,
the  resulting (average) pulse velocity $v = (\partial_T x_R + \partial_T
x_L)/2$
will also be small. Here $T=\eta^2 t$ is a slow time scale and $x_R$ and
$x_L$ are the position of the front at the right and at the left end of the
pulse, respectively.
Finally, in order to obtain a manageable expression for the solution of
(\ref{e:cac}) we assume
\be
\delta = \eta^4 \delta_4, \ \ \ \alpha =  \eta^2 \alpha_2, \ \ \ h = \eta^3
h_3, \ \ \
g = \eta^2 g_2.
\ee
This amounts to assuming the concentration mode to be small and its diffusion
and decay to be
 slow as compared to the velocity of
the front. These scalings lead to
\bea
\partial_tA+\eta^2 s_2 \partial_xA&=&\eta^2\partial_x^2A -A+cA^3-A^5+
CA,\label{e:abl}\\
\partial_tC&=&\eta^4 \delta_4\partial_x^2C-\eta^2\alpha_2 C+
\eta^3 h_3\partial_xA^2 + \eta^2 g_2 C A^2 .\label{e:cbl}
\eea
Due to the smallness of the diffusion coefficient of $A$
internal layers arise at the transitions from the conductive to
the convective state. Thus, in the presence of two fronts
the system has to be separated
into five regions. This is sketched in fig.\ref{f:regions}. It shows also
the typical shape of the fronts in the limit considered here.
Whereas the traveling-wave amplitude
is constant to lowest order in regions I, III and V, it changes rapidly from
the conductive to the convective state in the transition regions II and IV.
More
precisely, in II and IV $A$ and therefore also
$C$ varies on a fast space scale
$y = \eta^{-1} x$. In order to match the solutions in the different regions
the width $2\Delta$ of regions II and IV needs to be small on the scale $x$
but large on the scale $y$. This is achieved by taking
$\Delta = \eta^{1/2} \Delta_{1/2}$.

The amplitudes $A$ and $C$ as well as the control parameter $c$ are now
expanded as
\be
A=A_0+\eta A_1+...,\quad C=\eta C_1+..., \quad c=c_0 + \eta c_1.
\label{e:cadev}
\ee
Here  $c$  is expanded around the value $c_0\equiv4/\sqrt{3}$  for which in the
absence
of the concentration mode the conductive and
the convective state coexist, i.e. neither state invades the other.
In the transition regions II and IV one obtains then at $O(1)$
\be
0=\partial_y^2A_0+A_0+c_0A_0^3-A_0^5.\label{e:A0II}
\ee
At $O(\eta)$ and $O(\eta^2)$ one finds, respectively,
\bea
\partial_{T_1} A_0 + s_2 \partial_y A_0&=&\left(\partial_y^2+{\cal
L}\right)A_1+
C_1 A_0+c_1 A_0^3,\label{e:A1II}\\
\partial_{T_1} C_1&=&h_3\partial_yA_0^2.\label{e:C2II}
\eea
Here the intermediate time scale $T_1 = \eta t$ and the linearized operator
${\cal L}=1+3 c_0 A_0^2 - 5 A_0^4$ have been introduced.
Eq.(\ref{e:A0II}) yields $A_0(y)=A_f((x_R-x)/\eta)$ in IV and
$A_0(y)=A_f((x-x_L)/\eta)$ in II where the front solution is given by
$A_f(\zeta) = A^*\left(\frac{1}{2}(1+\tanh \zeta)\right)^{1/2}$ and
$A^{*2}=\sqrt{3}$.
Its position is given by $x_R$ in IV and by $x_L$ in II. Both
can vary slowly in time, $x_{R,L}=x_{R,L}(T)$.
Due to translational symmetry, the linearized
operator $\partial_y^2+{\cal L}$ is singular.
Therefore eq.(\ref{e:A1II}) leads to a solvability condition in each of the
two regions. Thus, by projecting eq.(\ref{e:A1II})
onto the left zero-eigenvector $\partial_yA_0$ one obtains evolution
equations for $x_R$ and for $x_L$, respectively. Note that in this approach
the distance between the two fronts does not have to be assumed to be large
in order to obtain two independent conditions. This is due to the small
diffusion of $A$; any finite distance on the scale $x$ is large on the
relevant scale $y$.

Since $A_0$ depends on time only via $x_{R,L}(T)$ the general solution
of eq.(\ref{e:C2II}) in II and IV is given by
\be
C_1=k_1^{II,IV}(T)-\frac {h_3}{\partial_T x_{R,L}} A_0^2
\ee
with $k_1^{II,IV}(T)$ being integration constants in the respective
regions. Note that this solution is only valid as long as $\partial_T x_{R,L}$
is $O(1)$ or larger. For smaller velocities and in particular for
stationary pulses the solution of the concentration equation becomes
more complicated, in particular when the diffusion of $C$ becomes relevant.

In regions I, III and V the traveling-wave amplitude
$A$ is constant to lowest order. Therefore, the source term $\partial_x A^2_0$
in the equation
for $C$ becomes less important than the damping term $-\alpha C$. Thus, in
these regions
eqs.(\ref{e:abl},\ref{e:cbl}) lead at $O(1)$, $O(\eta)$ and $O(\eta^3)$,
respectively, to
\bea
0=A_0+c_0A_0^3-A_0^5,\label{e:A0I}\\
\partial_{T_1} A_0={\cal L}A_1+C_1A_0+c_1A_0^3,\label{e:A1I}\\
\partial_{T} C_1=-\hat{\alpha}C_1\label{e:C4I}
\eea
where we have introduced the renormalized damping coefficient
\be
\hat{\alpha} \equiv \alpha_2 -g_2 A_0^2. \label{e:alphahat}
\ee
Note that for free-slip-permeable boundary conditions one obtains $g_2 > 0$
\cite{Ri92a},
i.e. the damping is reduced by convection. The solution of eq.(\ref{e:A0I}) is
given by
the extremal values of the front solution $A_f$.
Since ${\cal L}$ is non-zero in the regions in question, eq.(\ref{e:A1I}) can
always be solved for $A_1$ and no solvability conditions arise in
these regions.
Eq.(\ref{e:C4I}) has the general solution
\be
C_1=\tilde {C}_1(x)e^{-\hat{\alpha} T}
\ee
with the amplitude $\tilde{C}_1(x)$ as yet undetermined. Note that in
the present scaling the concentration mode decays only due to damping;
the diffusive term is irrelevant despite the fact that $C$ can vary on O(1)
length scales.

To make use of the solvability conditions the integration constants
$k_1^{II,IV}$ appearing in the solution for $C_1$ in II and IV are needed.
They are obtained by matching the solutions in the different regions.
In the limit of weak diffusion considered here ($\delta \ll 1$) the solution
depends
qualitatively on the direction of propagation of the two fronts.
In the following we concentrate on the case in which the two fronts travel in
the same direction and can therefore form a bound pair.
If they travel towards each other no interesting interaction arises
in this limit. For fronts traveling away from each other
the calculation would have to be modified. For concreteness we
assume both $v_R \equiv \partial_T x_R$ and $v_L \equiv \partial_T x_L$ to be
positive.

The traveling-wave amplitude vanishes in I and V and in III one has $A_0=A^*$.
Due to the damping term $\alpha C$, the concentration mode vanishes also far
ahead of the pulse, i.e. in V,
. This implies $k_1^{IV}=0$. At the right end
of region III $C_1$ is therefore given by
\be
C_1(x_R-\Delta,T) = - \frac{h_3}{v_R(T)} A^{*^2}.
\ee
To obtain $C_1$ at the left end of region III at time $t$, i.e. at
$x_L(T)+\Delta$, one has to
take into account that the concentration field at that location has been
generated by the leading edge of the pulse (which is now at $x_R(T)$) at
an {\it earlier} time $T^r \equiv T-\Delta T$ and has suffered an exponential
decay since then. Thus one obtains
\be
C_1(x_L+\Delta,T) = - \frac{h_3}{\partial_Tx_R(T-\Delta T)} A^{*^2}
e^{-\hat{\alpha} \Delta T},
\ee
where the delay $\Delta T$ is determined by
\be
x_R(T-\Delta T) = x_L(T).
\ee
Thus, in region II the concentration field $C_1$ is given by
\be
C_1 = h_3 A^{*^2} \left\{ \frac{1}{v_L(T)} (1-\frac{A_0^2}{A^{*^2}}) -
\frac{1}{v_R(T-\Delta T)} e^{-\hat{\alpha} \Delta T} \right\}.
\ee
For the solvability conditions $C_1$ need not be determined in region I.

Projecting eq.(\ref{e:A1II}) onto $\partial_y A_0$ leads in region
II and IV, respectively, to the equations
\bea
v_L(T) &=& s_2 - \frac{\gamma}{v_L(T)} +
\frac{2\gamma e^{-\hat{\alpha} \Delta T}}{v_R(T-\Delta T)}
- \rho,\label{e:vLdel}\\
v_R(T) &= &s_2 - \frac{\gamma}{v_R(T)} + \rho,\label{e:vRdel}\\
\partial_T L &=& v_R(T) - v_L(T)\label{e:Lvrvldel}
\eea
with $L=x_R-x_L$ and
\be
\gamma=\sqrt{3}h_3, \ \ \ \rho =\sqrt{3} c_1.
\ee
Here we have used
\be
\int_{-\infty}^\infty \, \left( \partial_y A_0 \right)^2
dy=A^{*2}/4=\sqrt{3}/4.
\ee
In terms of the front velocities $v_{R,L}(t)$ the delay $\Delta T$ is
given by
\be
\int_{T-\Delta T}^T\, v_R(T') dT'\, = L(T).\label{e:delay}
\ee
Thus, in the general case, in which $v_R$ depends on time,
eqs.(\ref{e:vLdel},\ref{e:vRdel})
with (\ref{e:delay}) represent a set of implicit integral-delay equations.
In the present scaling of weak diffusion, however, the dynamics in region IV is
decoupled
from that in region II. Therefore $v_R$ is constant and the retarded
velocity $v_R(T-\Delta T)$ is equal to $v_R$. Consequently, the delay is
given by $\Delta T = L/v_R$ and one obtains
\bea
v_R &= &s_2 - \frac{\gamma}{v_R} + \rho,\label{e:vR}\\
v_L(T) &=& s_2 - \frac{\gamma}{v_L(T)} +\frac{2\gamma e^{-\hat{\alpha}
L/v_R}}{v_R}
- \rho,\label{e:vL}\\
\partial_T L &=& v_R - v_L(T).\label{e:Lvrvl}\label{e:vr}
\eea
The results for negative front velocities are obtained in a similar way. Both
can be combined into equations for the velocities $v_{l,t}$ of the leading
and the trailing front, respectively,
\bea
v_l &=& s_2 - \frac{\gamma}{|v_l|} + sgn(v_l) \rho \label{e:vl}\\
v_t &=& s_2-\frac{\gamma}{|v_t|} +
2\gamma \frac{e^{-\hat{\alpha} L/|v_l|}}{|v_l|} - sgn(v_t) \rho,\label{e:vt}\\
\partial_T L &=& \gamma \left(\frac{1}{v_t}-\frac{1}{v_l}\right)
-  2\gamma \frac{e^{-\hat{\alpha} L/|v_l|}}{v_l} +2 \rho. \label{e:Lvlvt}
\eea
Here we have used $\partial_T L = sgn(v_l) (v_l-v_t)$ which assumes that both
fronts
travel in the same direction. Note that eq.(\ref{e:Lvlvt}) does not represent a
closed equation
for $L(T)$ due to the dependence of $v_t$ on $L$. The equation for $v_t$ is
quadratic in
$v_t$ and contains in addition the sign of $v_t$, $sgn(v_t)$.
Elimination of $v_t$ in favor of $L$ would therefore lead to
 complicated expressions involving
the exponential within square roots. We therefore prefer to discuss
eqs.(\ref{e:vl},\ref{e:vt}) for the velocities instead of the equation for
$L(T)$.

It should be pointed out that the validity of eqs.(\ref{e:vt},\ref{e:Lvlvt}) is
not limited to large values of $L$. Therefore $\gamma e^{-\hat{\alpha}
L/|v_l|}$ can be of the same order as $\gamma$ itself.
In the present scaling the traveling-wave amplitudes $A$ of the two fronts do
not overlap on this length
scale and do therefore not contribute to their interaction.
We will comment on the possible consequences of an overlap in $A$ and of
 dispersion at the end of sec.V.

\section{Discussion of the Front Equations}

In the limit $L \rightarrow \infty$ eqs.(\ref{e:vl},\ref{e:vt}) describe
 the dynamics of a single front.
 More precisely, a single front connecting the state $A=A^*$ for $x \rightarrow
-\infty$
with the state $A=0$ for $x \rightarrow +\infty$ corresponds in the present
calculation to
a leading front for $v>0$ (cf. fig.\ref{f:fronts}a) and to a trailing front for
$v<0$
(cf. fig.\ref{f:fronts}b). Its velocity $v$
is therefore given by the relationship
\be
\rho = v + \frac{\gamma}{|v|} -s_2 \label{e:rhov}
\ee
or equivalently by
\be
v=\frac{1}{2} \left( s_2+\rho \pm
\sqrt{(s_2+ \rho)^2-4sgn(v)\gamma} \right).\label{e:vrho}
\ee
Fig.\ref{f:vfront} gives a typical behavior of $v$
 as a function of $\rho$ for such a front. It clearly exhibits a
hysteretic transition from
a `fast' front to a `slow' front when decreasing $\rho$ from positive
 values through the saddle-node bifurcation at
\be
\rho_{SN}=2\frac{\gamma}{\sqrt{|\gamma|}}-s_2.
\ee
It is expected that only fronts on the top and on the bottom branch
are stable
and those on the middle branch are unstable. Since the main difference
between the fast and the slow front is
the strength of the concentration mode the destabilizing mode of fronts on the
middle branch
will mainly involve the concentration mode.

The possibility of a hysteretic
transition between fast and slow fronts is not unexpected. The source of the
concentration field is
given by
the gradients of the traveling front in the traveling-wave amplitude. If the
front is
traveling fast, the source remains at any given location only
briefly and therefore generates only a weak concentration field which
in turn has only little effect on the propagation velocity of the front. If the
front
is slow, however, a strong concentration field can be built up which
slows down the front considerably providing a positive feed-back mechanism.
Note that, in experiments, increasing $\rho$ corresponds to increasing
 the Rayleigh number. Thus, in principle it may be possible to
observe this drastic manifestation of the concentration mode experimentally.

A right-traveling front connecting $A=A^*$ for $x \rightarrow -\infty$ with
$A=0$ for $x \rightarrow +\infty$ is not
equivalent to one connecting $A=0$ with $A^*$ traveling to the left
(cf. fig.\ref{f:fronts}c).
This is due to the fact that $A$ is the amplitude of right-traveling
convection rolls. Reflection in space therefore yields left-traveling
rolls with opposite group velocity $s$ and
opposite advection of the concentration field, $h\partial_x A^2$.
The velocity of such a front is obtained from eq.(\ref{e:vrho})
by flipping the sign of $\rho$. The velocities for both types of fronts
are shown in fig.\ref{f:vtl}. In contrast to fig.\ref{f:vfront}
the velocity of the leading front is now
marked by heavy solid lines and that of the trailing front
-- in the limit $L \rightarrow \infty$ -- by
thin dashed lines.
For $\rho =0$ the convective and the
conductive state have the same energy in this dispersionless case. In the
absence of
 any interaction the trailing and
the leading front have therefore the same velocity. Due to reflection symmetry
the concentration field affects the velocity of both types of fronts in the
same way.
Thus, $v_l(\rho)$ and $v_t (\rho, L \rightarrow \infty)$ intersect at $\rho =0$
even in the presence
of the concentration field, as indicated by the three
open circles. These three solutions at infinite $L$
imply the existence of three different
localized waves (pulses) within this framework.

At finite distances the fronts interact with each other.
Fig.\ref{f:vtl} shows the somewhat complicated behavior arising from the
interaction.
The broken lines give the velocity $v_t$  of the trailing front
as a function of $\rho$ and with the velocity $v_l$ determined correspondingly.
The parameters are given by $\gamma=0.15$,
$\alpha=0.01$, $g=0$ and $s=1$. The distance between the fronts is $L=10$ in
 fig.\ref{f:vtl}a and $L=30$ in fig.\ref{f:vtl}b.
Eq.(\ref{e:vt})
is quadratic in $v_t$. One therefore obtains up to two solutions for given
$v_l$, $\rho$ and $L$, i.e.
the curve $v_t(\rho)$ can split up into two branches.
Eqs.(\ref{e:vl},\ref{e:vt})
are, however, only valid if $v_t$ and $v_l$ have the same sign.
Thus, for $v_t < 0$ only one branch
remains which is marked by heavy short dashes.
 Due to the interaction it is shifted to smaller values of $\rho$.
Consequently,
the intersection of $v_l$ and $v_t$, which signifies a pulse solution, occurs
for $\rho <0$.
For $v_t >0$ both branches are relevant. They are marked by heavy
dashed and dotted-dashed
lines, respectively. Due to the double-valued nature of $v_l$ one obtains four
different
values of $v_t$ for a given value of $\rho$, which are associated in a pairwise
manner
with the upper and the lower branch of $v_l$, respectively. Pulse solutions are
given by the
intersection points of the curves. Therefore,
care has to be taken which value of $v_t$ is associated with which branch of
$v_l$.
The relevant intersections are marked by solid circles.

The dependence of the velocity curves on the length $L$
allows conclusions about the character of the
interaction. For pulse $A$ in fig.\ref{f:vtl} it is repulsive;
without interaction (thin, dashed line) the magnitude of $v_t$ would be
larger than that of $v_l$ and the trailing front would catch up with the
leading
front. With decreasing distance $L$, however, the interaction strength
increases
and the magnitude of $v_t$ is reduced. Thus, the trailing front will approach
the leading front until their velocities match. If the distance were to be
decreased
further, the trailing front would start lagging behind until the equilibrium
distance
is reached again. This suggests that the repulsive interaction leads to a
stable pulse. By contrast, for pulse $B$ the interaction is attractive and
should
render it unstable: the closer the trailing front comes to the leading front
the faster it travels. For pulse $C$ the interaction is again repulsive.

 The above expectations about the stability of the pulses
are borne out in a direct linear stability analysis.
A linearization of eqs.(\ref{e:vl}-\ref{e:Lvlvt})
around the steady pulse solution with velocity $v_0$ and length $L_0$ (and
corresponding delay $\Delta T_0 = L_0/v_0$) yields for the growth rate $\sigma$
of perturbations
\be
\sigma = - \frac{2 \hat{\alpha} }{1-\frac{v_0|v_0|}{ \gamma}} e^{-\hat{\alpha}
L_0/|v_0|}.\label{e:sigma}
\ee
For $\hat{\alpha} >0$ the pulses are therefore stable as long as
$v_0|v_0|\gamma^{-1}<1$\footnote{Note that for $\hat{\alpha}<0$ the basic
equations (\ref{e:Atil},\ref{e:Ctil}) are not appropriate.}.
The saddle-node
bifurcation of the leading pulse occurs at  $v_0|v_0|=\gamma$. This
stability result
corresponds therefore to the two cases with repulsive interaction discussed
above
($A$ and $C$).
Only the case $v_0 <0$ will lead to stable pulses since on the lower branch
($0<v_0|v_0|<\gamma$) the leading
front itself is unstable to perturbations in the concentration field which take
it either
to the fast or to the slow branch. This instability is not contained in the
present
analysis of the velocity and interaction of fronts
(eqs.(\ref{e:vl},\ref{e:vt})),
since it involves a change in the character of
the front which is not merely given by a change in its position.

Can one understand the stability of backward traveling pulses intuitively?
Consider the experimentally relevant case $\gamma  >0$ \cite{Ri92a}.
Then the negative
gradient of $A^2$ at the right front leads to a negative peak of $C$ whereas
the left front generates a positive peak. If the pulse
is stationary, both peaks have the same magnitude due to reflection symmetry
as indicated in fig.\ref{f:regions} by the dotted line.
At the right front the concentration field reduces the local growth
rate  of $A$ and therefore pushes the pulse to the left. Similarly, the
peak at the left front enhances the growth rate und also pushes the pulse
to the left. Thus, in the stationary case both peaks change the
pulse velocity by the same amount and therefore do not contribute to the
interaction. If the pulse travels, however, this is not the case any more.
For negative velocities both peaks become steeper to their left side
and much shallower to the right (dashed line), and the concentration field of
the left
front influences the right front much more than the other way around.
Thus, while the left front enters a region with $C=0$, the right
front propagates into a region with $C>0$. The magnitude of the
concentration field at the
right front is therefore reduced as indicated by heavy lines in
fig.\ref{f:regions}. Consequently, the right front is pushed less to the left
than the left one resulting
in a repulsive interaction. This argument shows that
to lowest order the stability of the pulses requires
that their velocity $v$ be opposite to the group velocity
$s$: they travel {\it backwards}.

The dependence of the length of
the stable pulse on $\rho$ is shown in fig.\ref{f:Lrho}. As expected
the length diverges at $\rho=0$. It monotonically decreases with
increasing $|\rho|$. Strikingly, within this framework the stable pulse
exists for arbitrarily large negative values of $\rho$. To wit,
for large $\rho$ and small $v_l$ the velocity and length are given to leading
order by
\be
v_l=\gamma/\rho + ..., \ \ \ L=\frac{s_2}{\gamma \alpha} \frac{1}{\rho^2} +
..., \label{e:Lrho}
\ee
 implying that the repulsive interaction becomes arbitrarily large in this
limit. This is
due to the prefactor $1/v_l$ in eq.(\ref{e:vt}): with decreasing
magnitude of the velocity the build-up of the concentration field becomes
stronger and stronger. By contrast, for the (attractive)
interaction of fronts in the real Ginzburg-Landau equation alone
the prefactor of the exponential is constant and the strength of the
interaction remains
bounded.

\section{Numerical Results}

In the derivation of (\ref{e:vl},\ref{e:vt}) we have pushed the diffusion of
$A$
as well as the diffusive term in $C$ to higher order. To get an impression of
the influence
of these effects on single fronts and pulses we investigate
eqs.(\ref{e:caa},\ref{e:cac})
numerically. In fig.\ref{f:vfrontnum} the results for the velocity $v$ of a
single front
are given as a function of $c \equiv (4+\eta \rho)/\sqrt{3}$ for increasing
values
of $\delta$ \cite{HeRi94}. The large
symbols denote the velocity of a left-front, i.e. a front connecting
$A(x\rightarrow -\infty)=0$ with
$A(x\rightarrow +\infty)=A^*$, whereas the small open circles denote that of a
right-front.
These results show that the singularity in $c=c(v)$ at $v=0$ for a single front
disappears when the diffusive term for $C$ is included; instead the slow fronts
on the bottom
branch also undergo a saddle-node bifurcation and the two branches,
which are separated in the approximation leading to
eqs.(\ref{e:vl},\ref{e:vt}),
are connected in an S-shaped curve. With increasing $\delta$ the S-shape
becomes less pronounced and one obtains eventually a continuous transition to a
single-valued velocity.
A similar transition occurs with decreasing coupling strength.

The diffusive terms have also a strong effect on the regime of
existence of the stable pulse. The results of numerical simulations of
eq.(\ref{e:caa},\ref{e:cac}) are shown in fig.\ref{f:vLpulsnum}(a,b) which give
the length
$L$ and the velocity $v$ of the pulse as a function of $c$ for different values
of $\delta$.
While the length appears to diverge with increasing $c$, the pulses disappear
in a saddle-node
bifurcation at small values of $c$.

This behavior can be understood qualitatively in the
following way. In the presence of diffusion not only the trailing but also
the leading front is affected by the interaction. For very weak diffusion,
the effect on the leading front will, however,
be negligible. Nevertheless,
the solution branches cannot be separated into trailing and leading fronts
any more. Instead one has
to label them again as left and right fronts with velocities $v_L$ and $v_R$.
Considering the diffusionless result, we expect that for weak diffusion the
stability
of the pulses requires that their velocity be negative and that the
isolated left-front be slower than the right-front. Therefore,
$c$ must be larger than a minimal value $c_{min}$ at which $v_L$ vanishes and
below a
maximal value $c_{max}$ at which $v_L=v_R$. This is sketched in
fig.\ref{f:vlrsketch}.
The difference $\Delta v$ in the velocities has to be compensated by the
interaction term, which will now also depend on $\delta$. Keeping the main
ingredients of the interaction in
mind one may expect it to be of the form $f(v) e^{-\hat{\alpha}L/|v|}$ where
$f(v)$ remains
now bounded for $v \rightarrow 0$. Balancing the two terms one obtains the
relationship
$L = - \hat{\alpha}^{-1}|v|\ln \left(\Delta v/f(v)\right)$. Thus, if $\Delta v
\rightarrow 0$
at finite $v$ the length diverges. This is the case at $c_{max}$. If, however,
$v \rightarrow 0$ at finite $\Delta v$, as is the case at $c_{min}$, $L
\rightarrow 0$.

Of course, when $L$ reaches $O(\eta)$ the overlap in the traveling-wave
amplitude $A$
has to be taken into account, as well.
Since for very small lengths all pulse-like initial conditions will collapse,
a short, unstable pulse has to exist which separates the basins of attraction
of the
pulse-less solution ($A=0$) and that of the stable pulse. With decreasing $c$
the
stable pulse is therefore expected to collide with this unstable pulse
in a saddle-node bifurcation for $c$ slightly above $c_{min}$. The
unstable pulse should correspond to that of the uncoupled real
Ginzburg-Landau equation where it exists for $c > c_{max}$ and has diverging
length for $c \rightarrow c_{max}^+$. In the presence of the concentration mode
it appears to
be created together with the stable pulse already for $c < c_{max}$ at a finite
length.

With increasing $\delta$ - and also with decreasing $h$ - the hysteresis in the
front velocities decreases. The range $(c_{min},c_{max})$ will therefore
decrease
and vanish above some
critical value of $\delta$.  Thus, stable pulses exist only for sufficiently
strong
coupling $h$ and sufficiently weak diffusion $\delta$ (cf.
fig.\ref{f:vLpulsnum})).

In principle, it is conceivable that the left-front solution reaches its
saddle-node bifurcation at negative $v_L$. In that case the branch of stable
pulses has to cease because no
suitable leading front exists. The length at which this occurs could be
sufficiently
large that the overlap in $A$ is irrelevant and may diverge with increasing
$\delta$ or
decreasing $h$.

\section{Conclusion}

In this paper we have investigated the dynamics of fronts
which interact via a long-wavelength mode. The work was motivated by convection
in
binary liquid mixtures where the fronts connect the motionless state with the
convective state and the long-wavelength mode corresponds to a concentration
mode which arises due to
small mass diffusion. We have shown that this
additional mode can lead to stable pulses in the form of
bound pairs of fronts even in the
absence of dispersion.

How do these results relate to experiments in binary-mixture convection?
Of course, the physical system is not dispersionless, i.e. the coefficients
in equation (\ref{e:Atil}) as well as the traveling-wave amplitude are in
general complex. It has
been shown previously \cite{MaNe90,HaPo91} that this by itself can lead
to stable bound pairs of fronts without the presence of a concentration
mode. For weak dispersion one obtains an equation of the
form \cite{MaNe90,HaPo91}
\be
\partial_T L = k_1 (c-c_0) - k_2 e^{-L/\xi} + \frac{k_3}{L} \label{e:Lvdisp}
\ee
for the length $L$ of a pulse with $k_1$ and $k_2$ positive.
The exponential term represents the attractive interaction between
fronts due to the overlap of their exponential tails in the convective
amplitude, while
the algebraic interaction term arises from dispersion. For suitable
parameters it can be repulsive ($k_3 >0$). Within this framework,
for $k_3 > 0$ and $c < c_0$ one
has two pulse solutions as in the dispersionless case discussed above.
Again the one with larger $L$ is the stable
one\footnote{In the regime in question eq.(\ref{e:Lvdisp}) has in fact three
solutions. The third solution, which is formally stable within
eq.(\ref{e:Lvdisp}),
is, however, always confined to $L/\xi=O(1)$. This small distance between the
fronts
is inconsistent with the assumption
of weak interaction between them since $k_2 =O(1)$. In fact, this solution is
unphysical;
as discussed in the previous section, the smallest pulse must be unstable.}.
This ordering appears to persist for strong dispersion where
the stable pulse may be identified with a dissipatively perturbed soliton
solution of the nonlinear Schr\"odinger equation \cite{HaJa90}.

In recent experiments,
both stable as well as unstable pulses have been investigated in great detail
\cite{Ko94}. The unstable ones have been obtained by employing a
suitable servo control. In that work two regimes have been found.
For a value of the separation ratio of $\psi= -0.127$ two
pulses exist for a given Rayleigh number and the stable
pulse is the {\it shorter} one. For $\psi \le -0.17$ the branch with the
longer, unstable pulses turns around and one is left with a
single pulse solution which is stable.

It is tempting to speculate that the experimental finding of longer,
unstable pulses can be described by including dispersion as well as
the concentration mode. The derivation of the relevant equations appears
to be involved. Considering eqs.(\ref{e:Lvdisp})
and (\ref{e:vl},\ref{e:vt},\ref{e:Lvlvt}) one may expect that
an equation of the form
\be
\partial_T L = k_1 (c-c_0) - k_2 e^{-L/\xi} + \frac{k_3}{L} -
2\frac{\gamma}{v_l}e^{-\hat{\alpha} L/|v_l|}\label{e:Lvall}
\ee
could capture the main aspects. With $\gamma > 0$ as before
 this equation allows
four solutions of forward traveling pulses ($v_l >0$). This is illustrated
in fig.\ref{f:Lvall} where the $dL/dt$ is plotted as a function of $L$ for
two exemplary cases.  Again the shortest pulse at $L=0.25$ (marked by an open
triangle)
is unphysical. The unstable pulse $A$ constitutes the
separatrix between the pulseless conductive state and the state with a stable
pulse ($B$). It has not been observed. We identify
 the experimentally observed pulses with the stable
pulse $B$ and the unstable pulse with pulse $C$. Thus,
within the extended equations an unstable longer pulse can coexist with a
shorter stable one as seen in the experiment.
Reducing $c$ slightly below $c_0$ leads
to a second stable pulse ($D$). Since its length diverges as $c \rightarrow c_0
$
it may be too long to be observed in the present experiments.

When $\gamma/v_l >0 $ is decreased towards $0$ the two pulses $B$ and $C$ merge
in a saddle-node bifurcation and - in particular for backward traveling pulses
($v<0$) -
one is left with the unstable pulse $A$ and the stable pulse $D$. We identify
this
situation with that found in the experiments for $\psi < -0.17$.
This interpretation is supported by the fact that in the experiments the
stability of the long pulses changes very close to where their velocity changes
sign.
Within the present scenario one would expect the length scale for
the long, stable pulses to be set by the decay
length $\ell \equiv |v|/\hat{\alpha_2}$ of the concentration field. Since
convection has a strong
influence on the decay rate through $\hat{\alpha} = \alpha_2 - g_2 |A_0|^2$,
such
an estimate requires more than the knowledge of the Lewis number and the
measured velocity
of the pulses. It is therefore not too surprising that taking $\alpha_2$ alone
instead of $\hat{\alpha}$
leads to an unreasonably short decay length $\ell$.
Clearly more detailed calculations are necessary to substantiate the discussion
based on
eq.(\ref{e:Lvall}) \cite{RaRi94c}.

In this paper we have concentrated on localized solutions which can be
considered as bound pairs of fronts, each of which forms a heteroclinic orbit
in space. This requires, of course, a subcritical
bifurcation. It is worth noting that numerical simulations of the extended
Ginzburg-Landau
equations (\ref{e:caa},\ref{e:cac}) have shown that localized waves can also
arise in the case of a supercritical bifurcation, for which no single fronts
exist \cite{Ri92a}.
These pulses correspond to homoclinic orbits in space. So far, they have not
been
captured in any analytic description.

This work was supported by DOE through grant (DE-FG02-92ER14303) and by
an equipment grant from NSF (DMS-9304397).
H.H. was supported by a grant of the F.P.I. program Ref.AP90 09297081 (M.E.C.).
H.R. thanks the Universidad de Navarra for its hospitality.
We gratefully acknowledge helpful suggestions by W.L. Kath and discussions with
V. Hakim and with P. Kolodner, who kindly shared his data with us prior to
publication.


\begin{figure}
\caption{Influence of the propagation of the pulse on the concentration field.
A stationary
pulse has symmetric concentration field (dotted line for $C$). In a propagating
pulse (dashed
line for $C$) the trailing peak of $C$ is reduced leading to an effective
repulsion between the fronts (see text).
\protect{\label{f:regions}}
}
\end{figure}

\begin{figure}
\caption{Sketches of leading and trailing fronts for $v>0$ and $v<0$.
a) leading front with $v>0$. b) trailing front with $v<0$.
c) trailing front with $v>0$. d) leading front with $v<0$.
\protect{\label{f:fronts}}
}
\end{figure}

\begin{figure}
\caption{Dependence of the velocity of a single front on $\rho$.
\protect{\label{f:vfront}}
}
\end{figure}

\begin{figure}
\caption{Velocity of a leading (solid line) and of a trailing front (dashed
lines)
as a function of $\rho$. Thin dashed lines denote $v_t$ for $L \rightarrow
\infty$.
Thick dashed lines give $v_t$ for $L=10$ and $L=30$, respectively.
\protect{\label{f:vtl}}
}
\end{figure}

\begin{figure}
\caption{Dependence of the length of pulses on $\rho$.
\protect{\label{f:Lrho}}
}
\end{figure}

\begin{figure}
\caption{Numerical results for the velocity $v$ of a single front on $c$. The
parameter values
 are $s=0.3$, $\eta^2 =0.05$, $\alpha =0.01$ and $h=0.03$.
\protect{\label{f:vfrontnum}}
}
\end{figure}

\begin{figure}
\caption{Sketch of the influence of diffusion and damping of $C$ on the
velocities
$v_{l,t}$ of fronts. Solid lines for $L\rightarrow \infty$, dashed line gives
$v_t$
 for finite $L$.
\protect{\label{f:vlrsketch}}
}
\end{figure}

\begin{figure}
\caption{Numerical simulations of pulses with $s=0.3$, $\eta^2 =0.05$, $\alpha
=0.01$
and $h=0.03$. The grid size is $\Delta x =0.1$ and the time step is $\Delta t
=1$. The
simulations ar done in a frame moving with the pulse.
\protect{\label{f:vLpulsnum}}
}
\end{figure}

\begin{figure}
\caption{$dL/dt$ as a function of $L$ according to
eq.(\protect{\ref{e:Lvall}}).
In this regime eq.(\protect{\ref{e:Lvall}}) has up to 4 steady pulse solutions.
\protect{\label{f:Lvall}}
}
\end{figure}

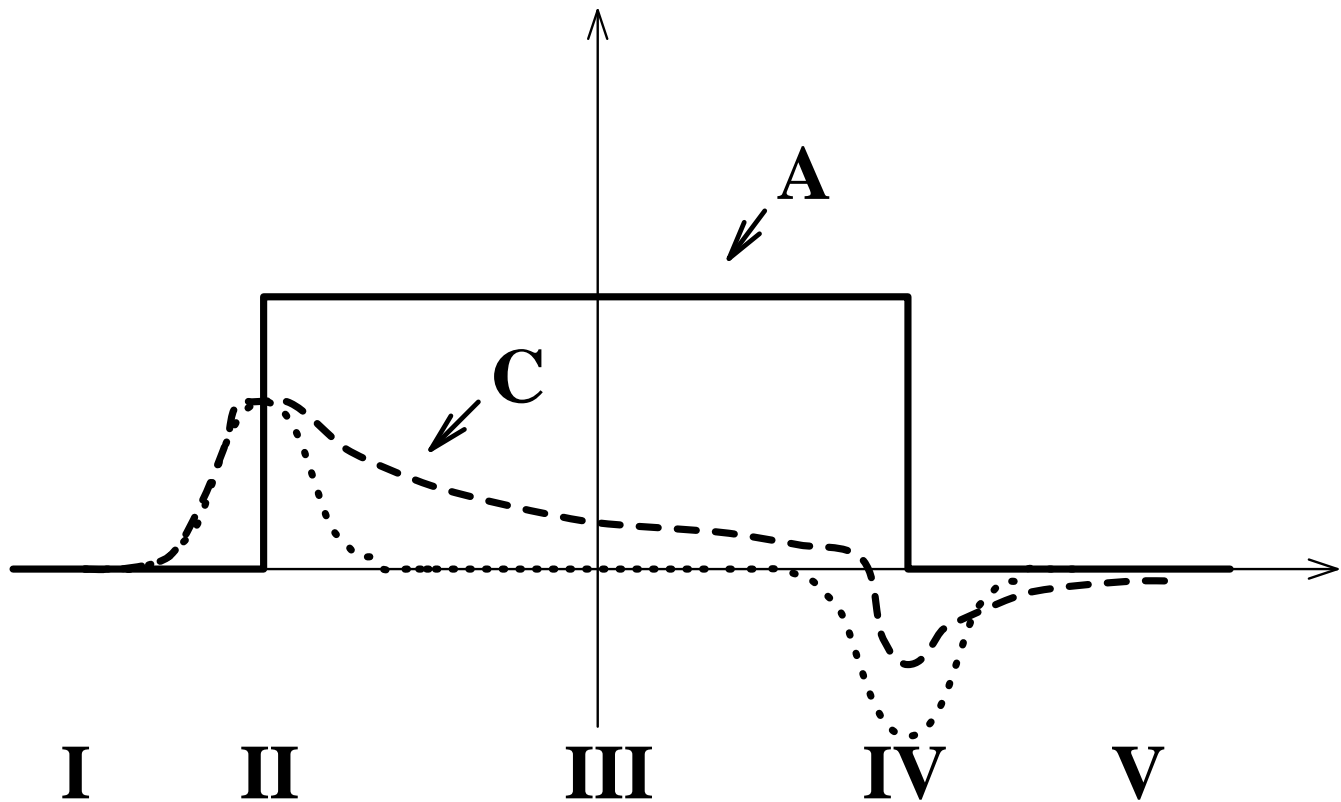
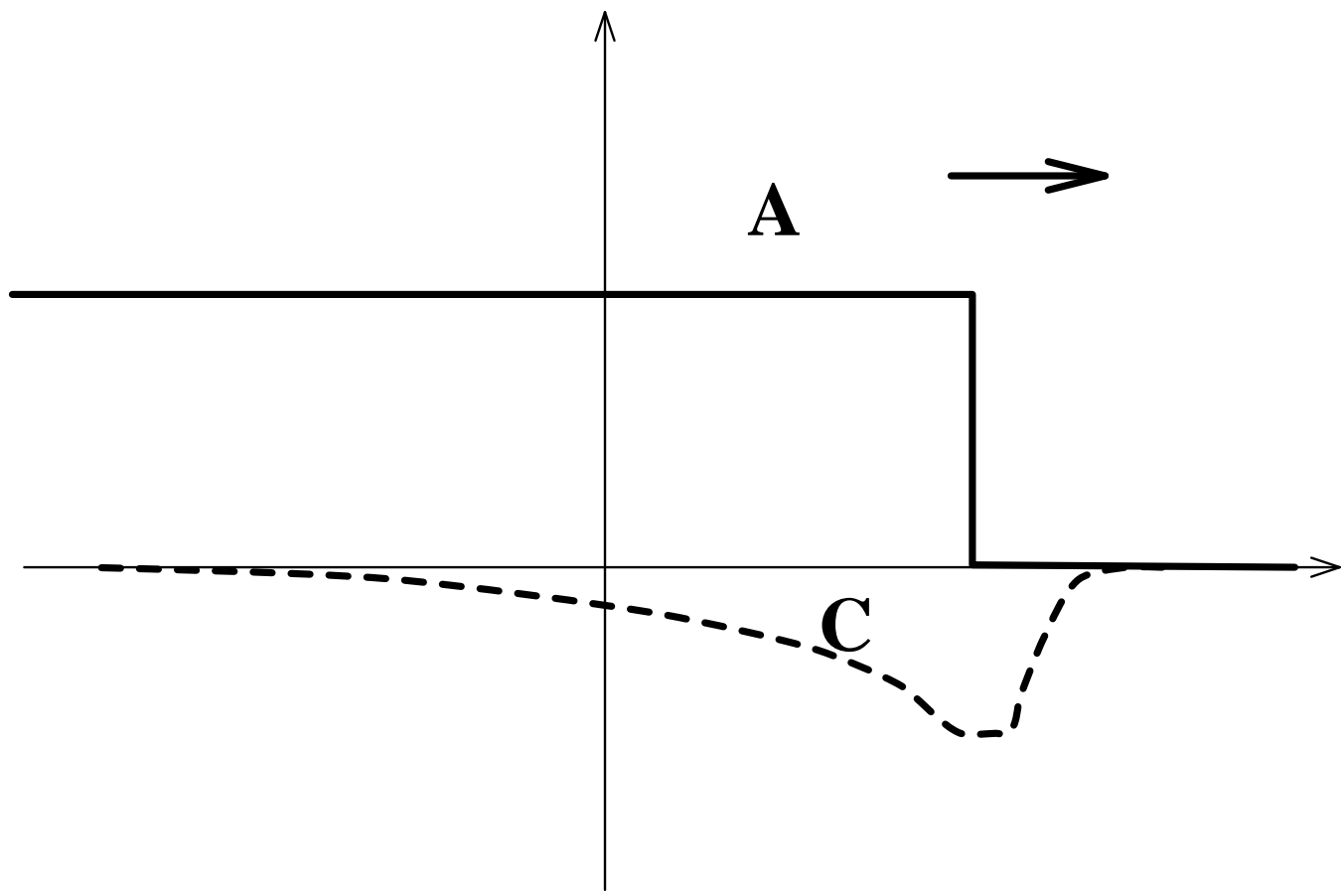
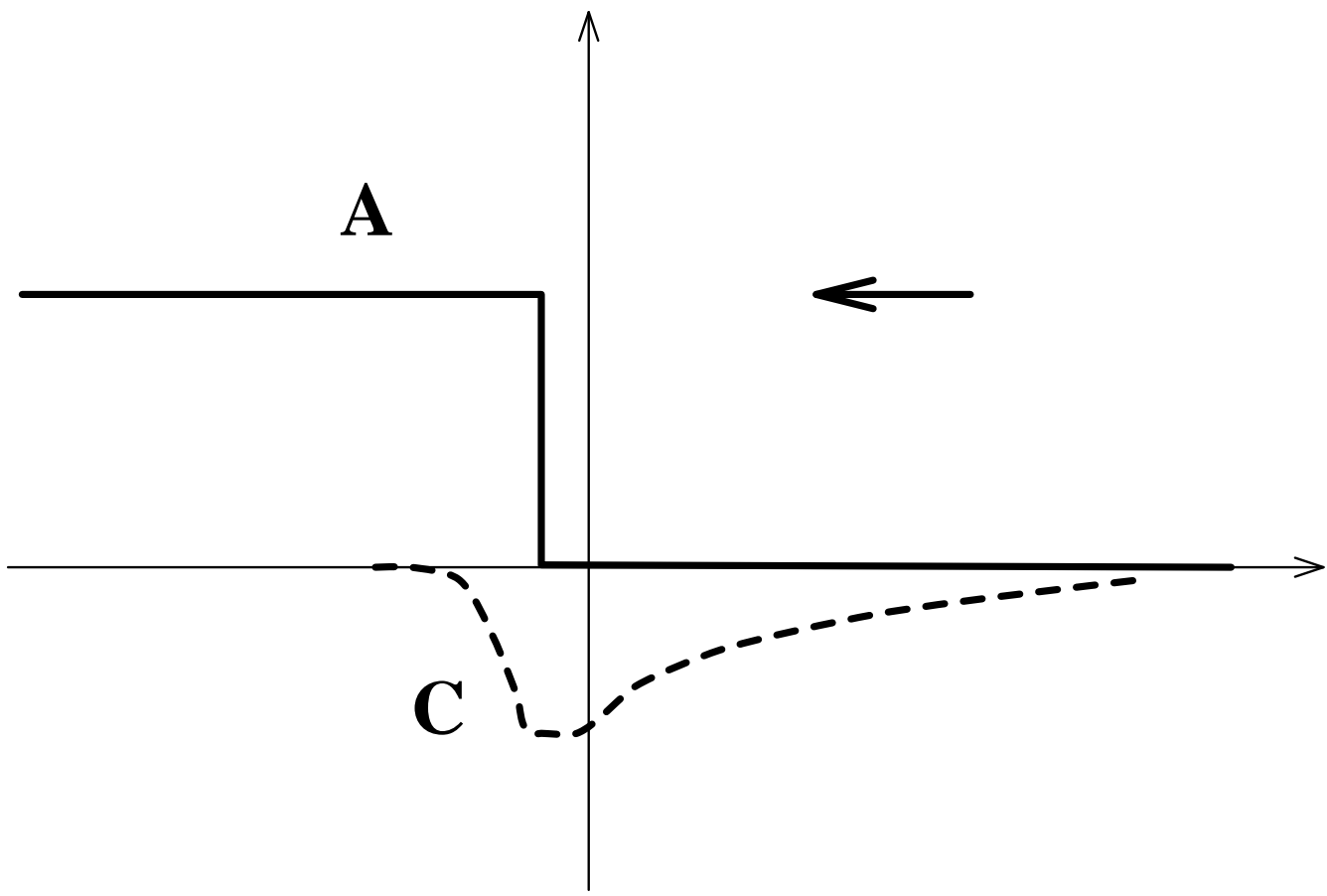
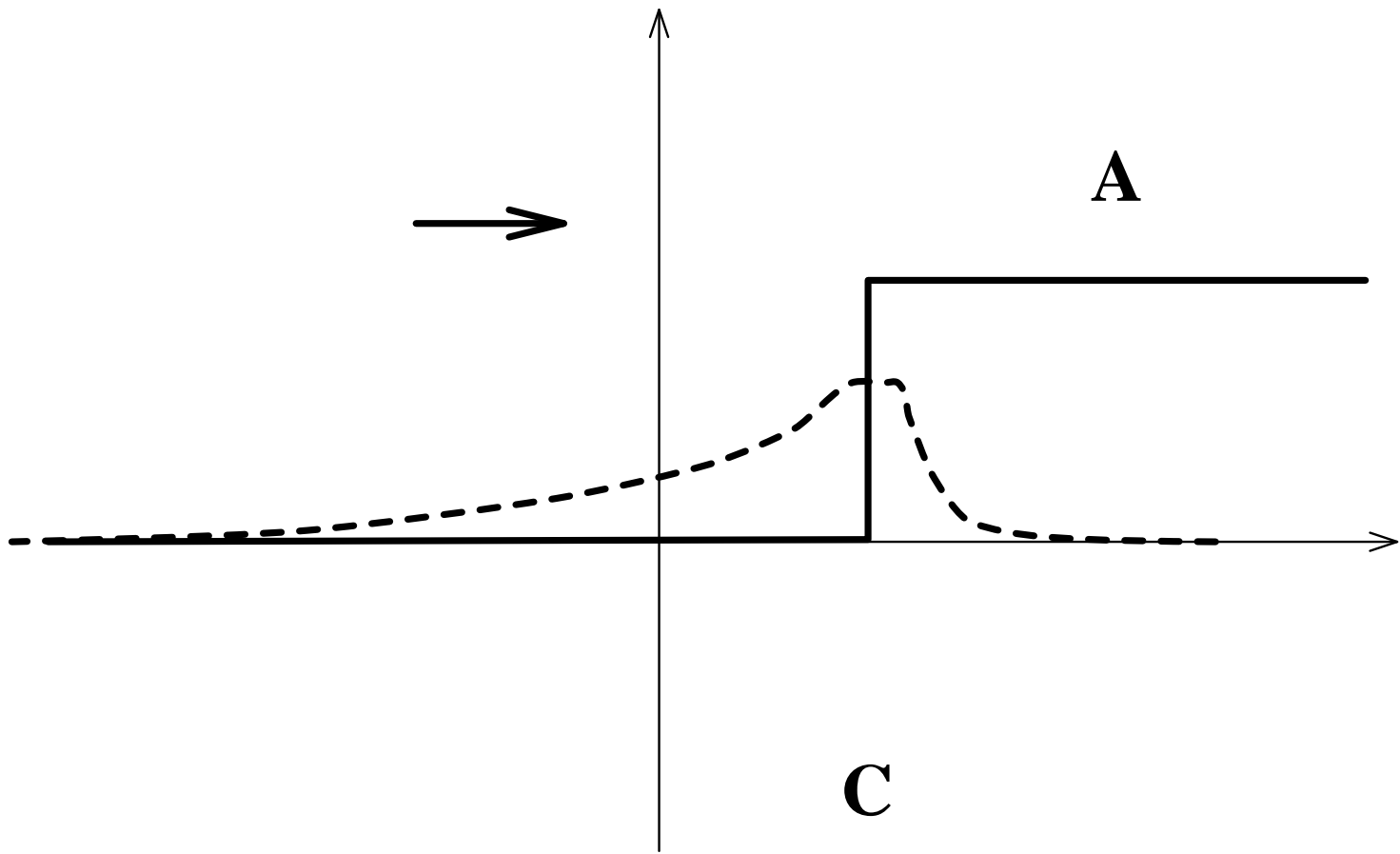
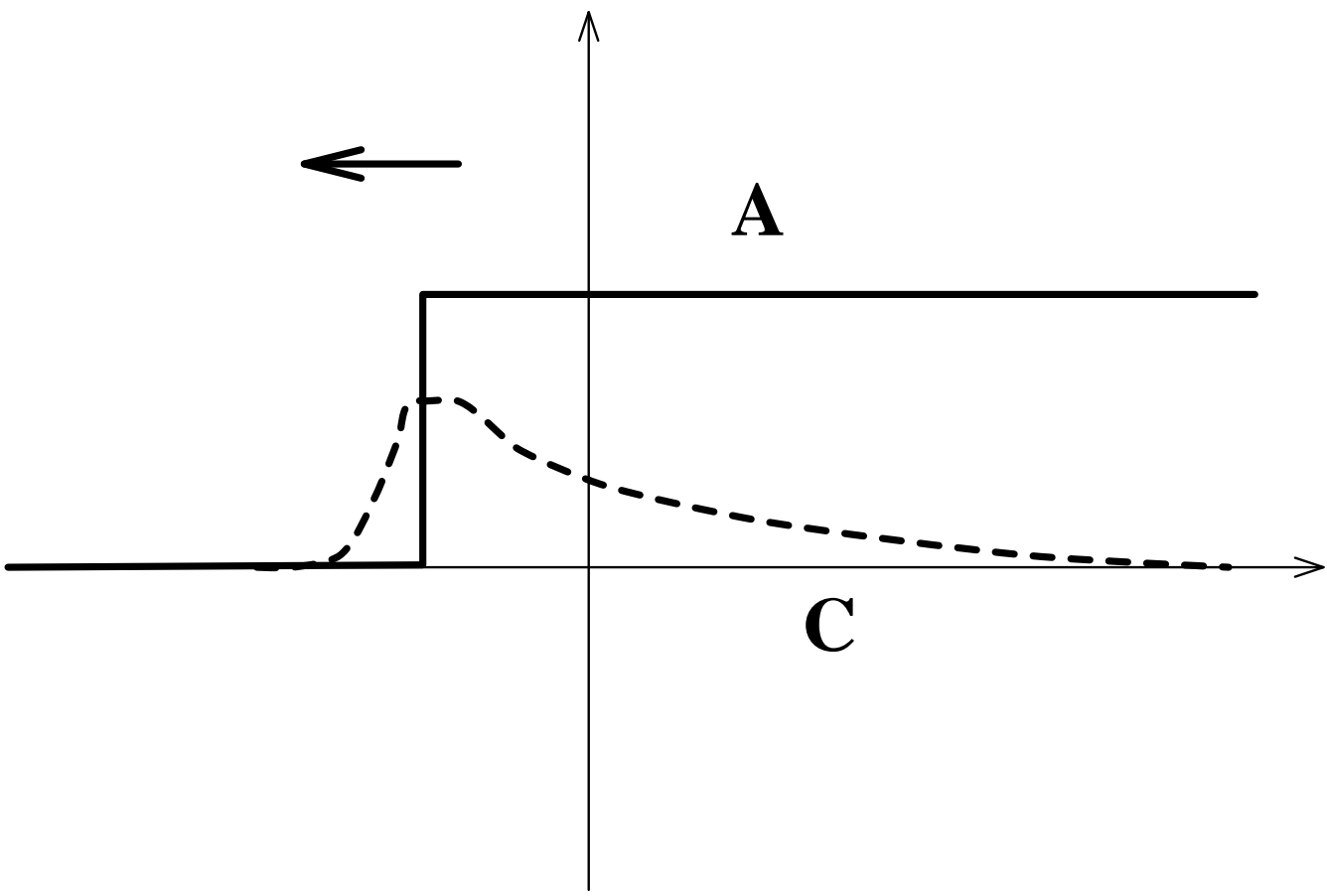
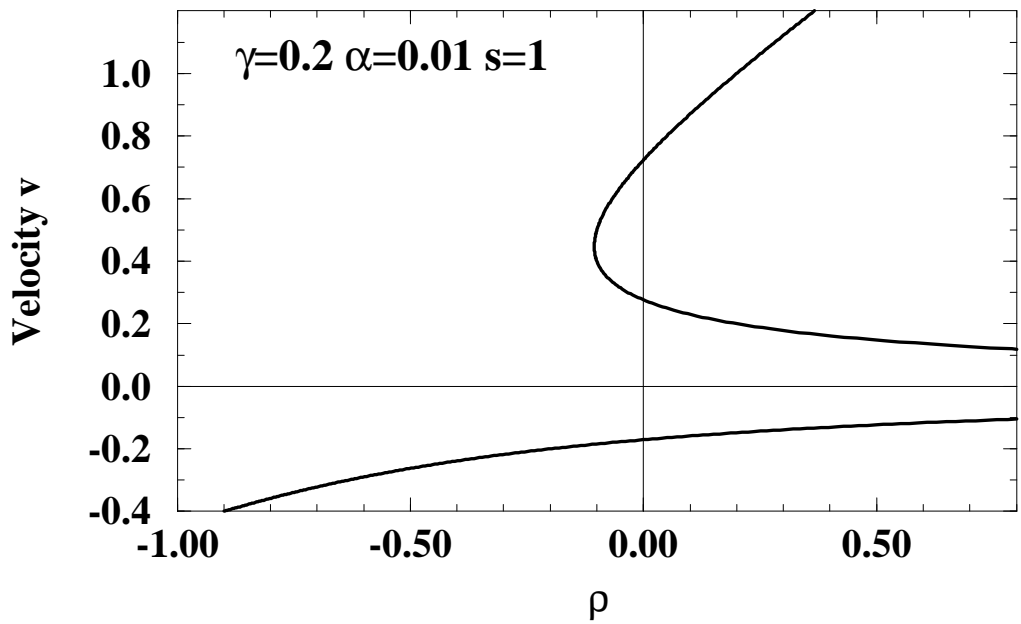
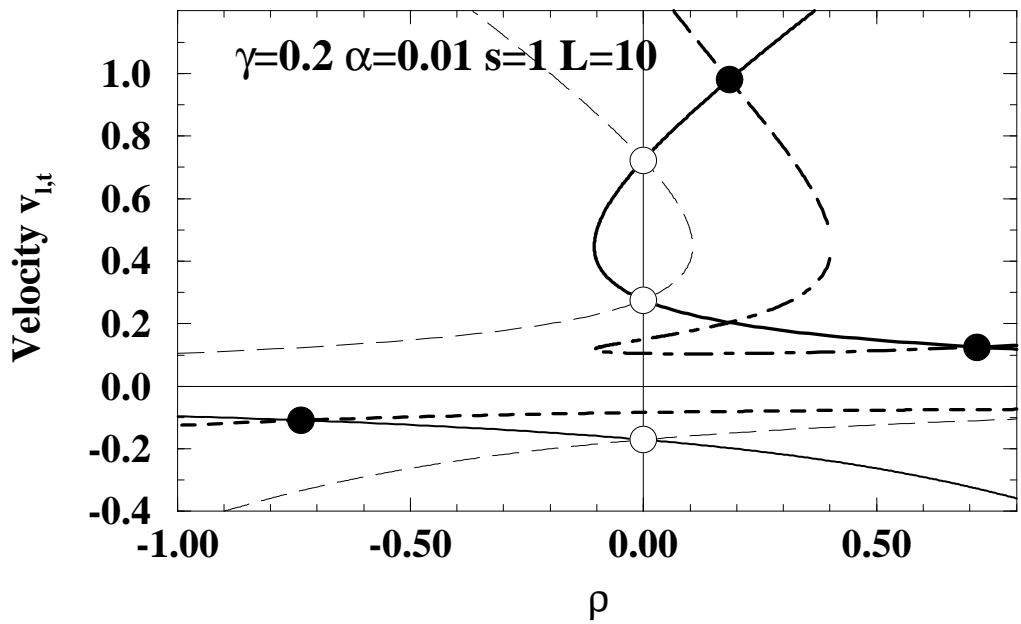
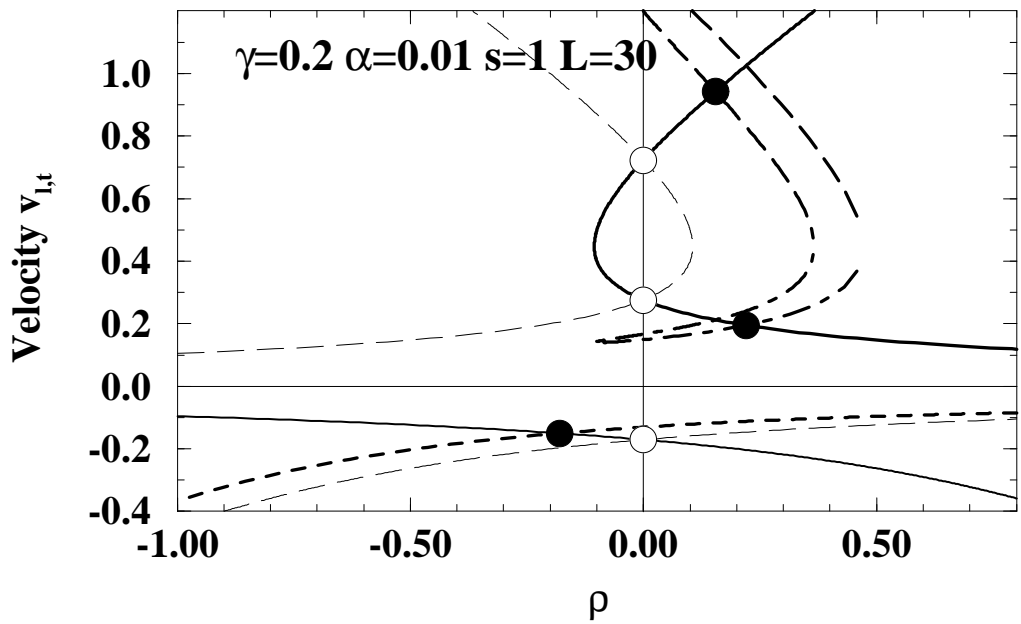
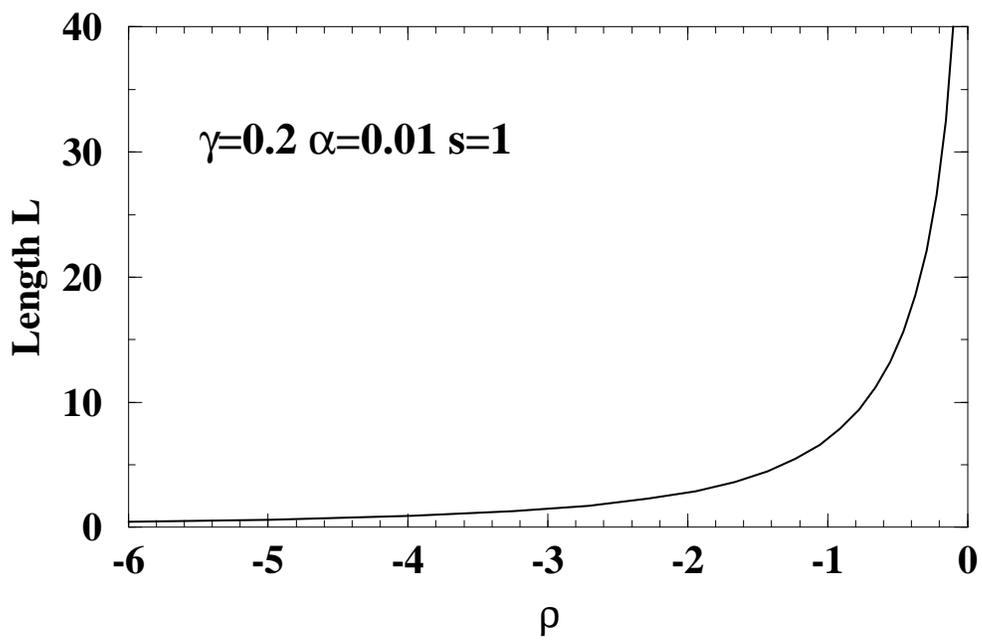
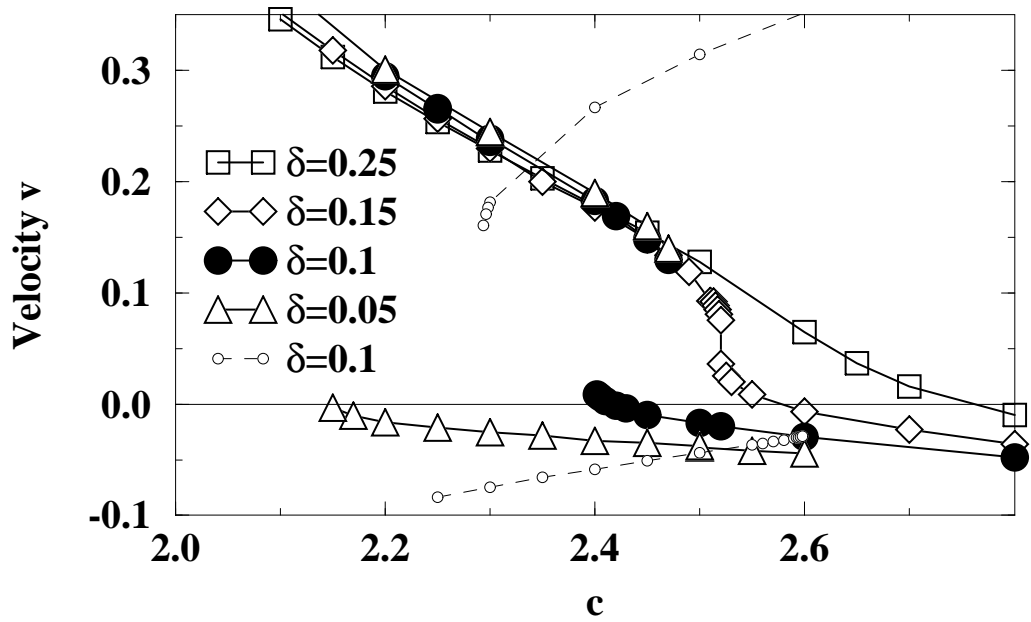
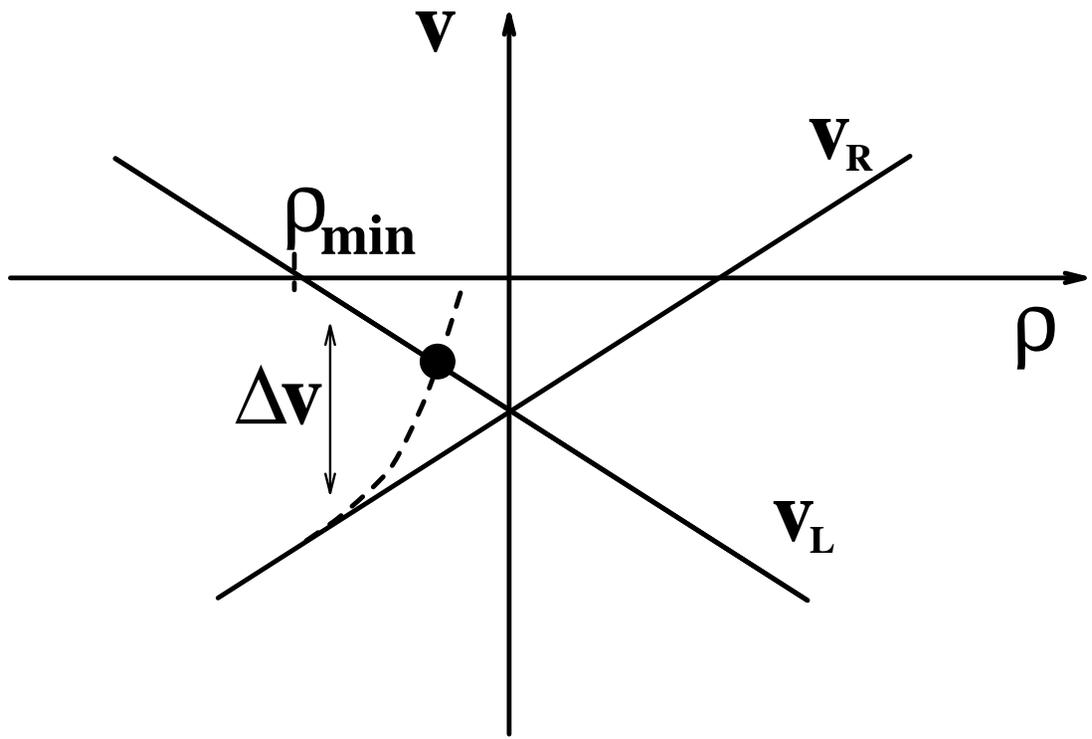
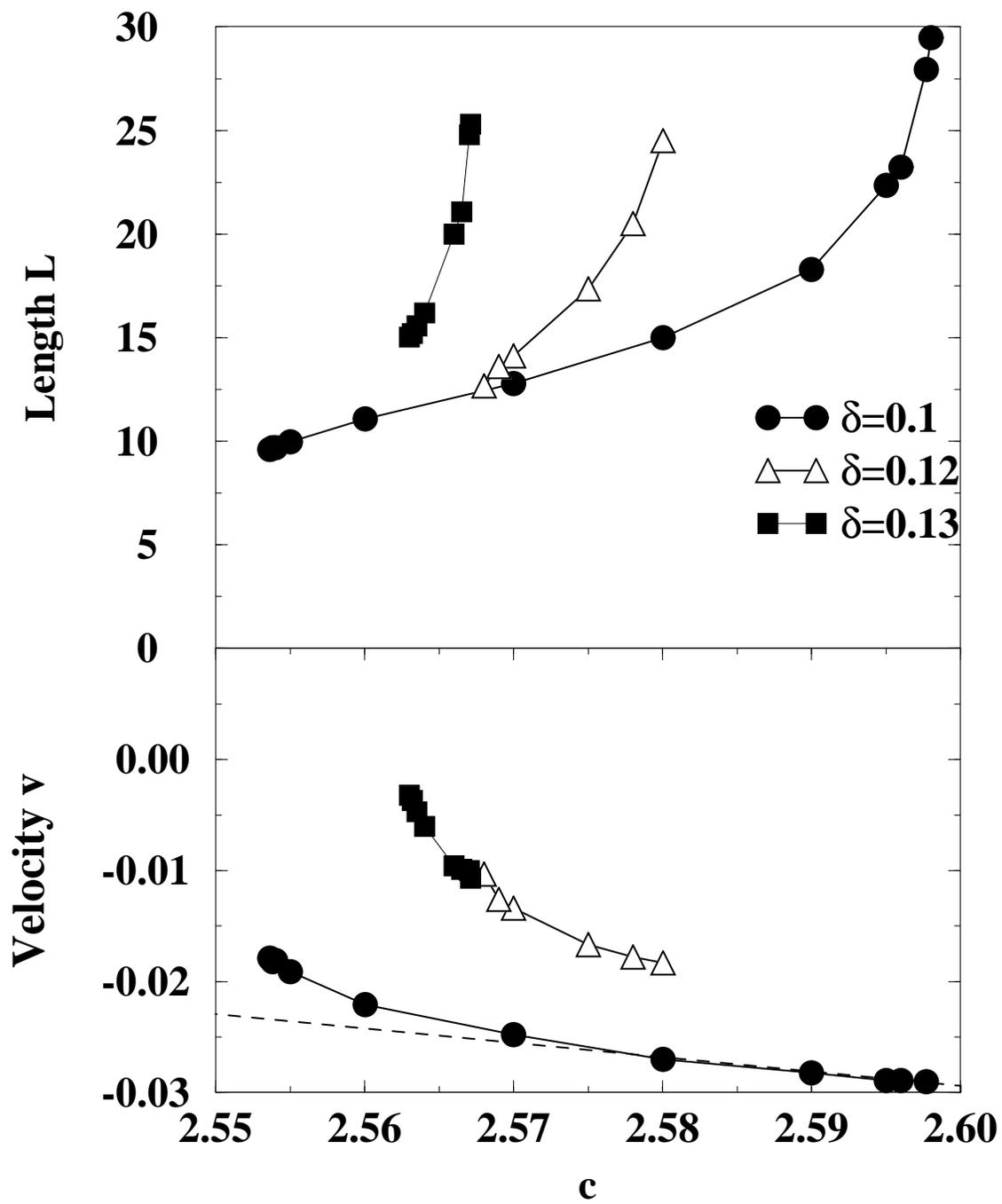
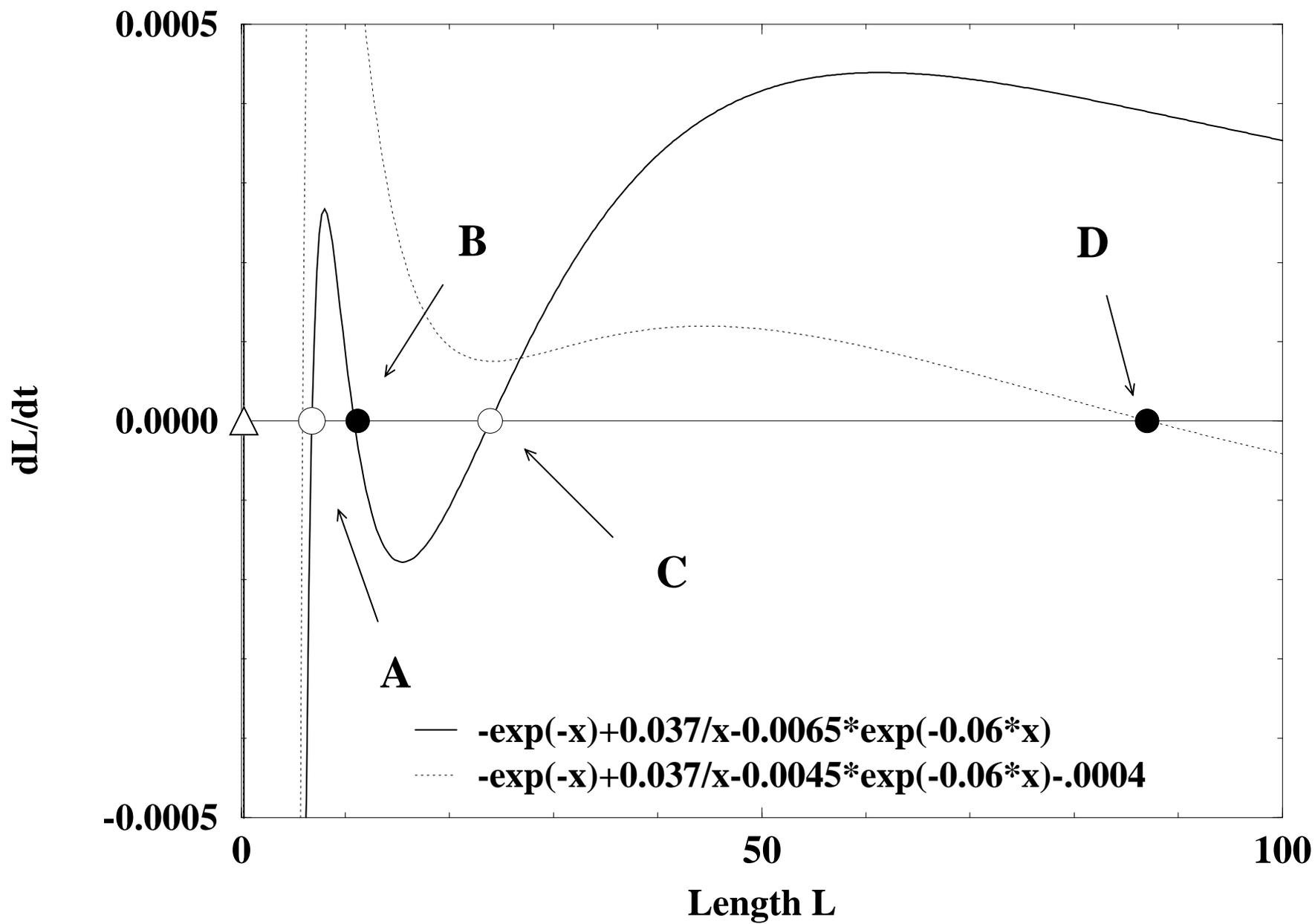
\end{document}